\documentclass{nature}

\usepackage{graphicx}
\usepackage{epsfig}

\title{Position-dependent diffusion of light \\
in disordered waveguides}

\author{Alexey G. Yamilov$^{1}$\footnote{e-mail:yamilov@mst.edu}, Raktim Sarma$^2$, Brandon Redding$^2$, Ben Payne$^{1}$, Heeso Noh$^{2,3}$ \& Hui Cao$^{2,4}$\footnote{e-mail:hui.cao@yale.edu}}

\begin{document}

\maketitle

\begin{affiliations}
 \item Department of Physics, Missouri University of Science \& Technology, Rolla, Missouri 65409, USA 
 \item Department of Applied Physics, Yale University, New Haven, Connecticut 06520, USA
 \item Department of Nano and Electronic Physics, Kookmin University, Seoul 136-702, Korea
 \item Department of Physics, Yale University, New Haven, Connecticut 06520, USA
\end{affiliations}

\begin{abstract}
Diffusion has been widely used to describe a random walk of particles or waves, and it requires
only one parameter -- the diffusion constant. For waves, however, diffusion is an approximation that disregards the possibility of interference\cite{1999_van_Rossum}. Anderson localization\cite{2009_Lagendijk_PT}, which manifests itself through a vanishing diffusion coefficient in an infinite system\cite{1980_Vollhardt_Wolfle,1993_Kroha_self_consistent}, originates from constructive interference of waves traveling in loop trajectories -- pairs of time-reversed paths returning to the same point\cite{1979_Gorkov,1981_Hikami}. In an open system of finite size, the return probability through such paths is reduced, particularly near the boundary where waves may escape. Based on this argument, the self-consistent theory of localization and the supersymmetric field theory predict\cite{2000_van_Tiggelen,2008_Cherroret,2008_Tian} that the diffusion coefficient varies spatially inside the system. A direct experimental observation of this effect is a challenge because it requires monitoring wave transport inside the system. Here, we fabricate two-dimensional photonic random media and probe position-dependent diffusion inside the sample from the third dimension. By varying the geometry of the system or the dissipation which also limits the size of loop trajectories, we are able to control the renormalization of the diffusion coefficient. This work shows the possibility of manipulating diffusion via the interplay of localization and dissipation. 
\end{abstract}

As first shown by Einstein in his theory of Brownian motion, the diffusion equation describes the evolution of the density of particles each undergoing a random walk\cite{1905_Einstein_Random_Walk}. The concept of diffusion has since been used to describe transport phenomena in both physics and other sciences. The power of this approach is that it requires knowledge of a single parameter, the diffusion constant, regardless of the underlying microscopical mechanisms of transport. If the spatial gradient of particle density is not too large, the particle flux is linearly proportional to the gradient, and the coefficient is the diffusion constant. Diffusion is also applicable to waves\cite{1960_Chandra,1978_Ishimaru,1980_van_de_hulst}, but it ignores interference effects. When inelastic scattering is negligible, most of the elastically scattered waves have uncorrelated phases and their interference is averaged out. Nevertheless, a wave may return to a position it has previously visited after a random walk, and there is always the time-reversed path which yields identical phase delay. Constructive interference of the waves from the reversed loops increases wave (energy) density at the original position and decreases the flux, giving the so-called weak localization effect\cite{1979_Gorkov}. This is the basic mechanism for the suppression of wave diffusion, which eventually leads to Anderson localization\cite{1958_Anderson}. 

In the self-consistent theory of localization, the diffusion coefficient $D$ is renormalized, and the amount of renormalization is proportional to the return probability of waves via the looped paths\cite{1980_Vollhardt_Wolfle,1993_Kroha_self_consistent}. In an open system of finite size, the return probability is reduced because the longer loops may reach the boundary where waves escape. Thus the renormalization of $D$ depends on the system size.  Moreover, near the boundary the chance of escape is higher, so the renormalization of $D$ is weaker. This means the value of $D$ is no longer constant but varies spatially\cite{2000_van_Tiggelen,2008_Cherroret,2008_Tian}. In the presence of dissipation the long loops are also cut, thus the renormalization of the diffusion coefficient depends on the amount of dissipation. This sets an effective system size beyond which the wave will not return\cite{2010_Payne_PRL}. Although self-consistent theory has successfully interpreted several experiments\cite{2006_Maret_PRL,2008_van_Tiggelen_Nature,2009_Genack_PRB,2013_Maret_3D_localization}, its key prediction of position-dependent diffusion has not been observed directly because it is difficult to probe wave transport inside the system experimentally. 

Here we report direct experimental evidence of position-dependent diffusion by probing light transport inside a quasi-two-dimensional random system from the third dimension. The system size and shape are designed to make the return probability sufficiently high so that the diffusion coefficient is modified appreciably. We also use dissipation to control the effective system size, and tune the value of $D$ via the interplay of localization and dissipation. This work demonstrates the possibility of utilizing the geometry of a random system or the dissipation to manipulate wave diffusion. 

\begin{figure}
\centering{\includegraphics[width=4in,angle=-0]{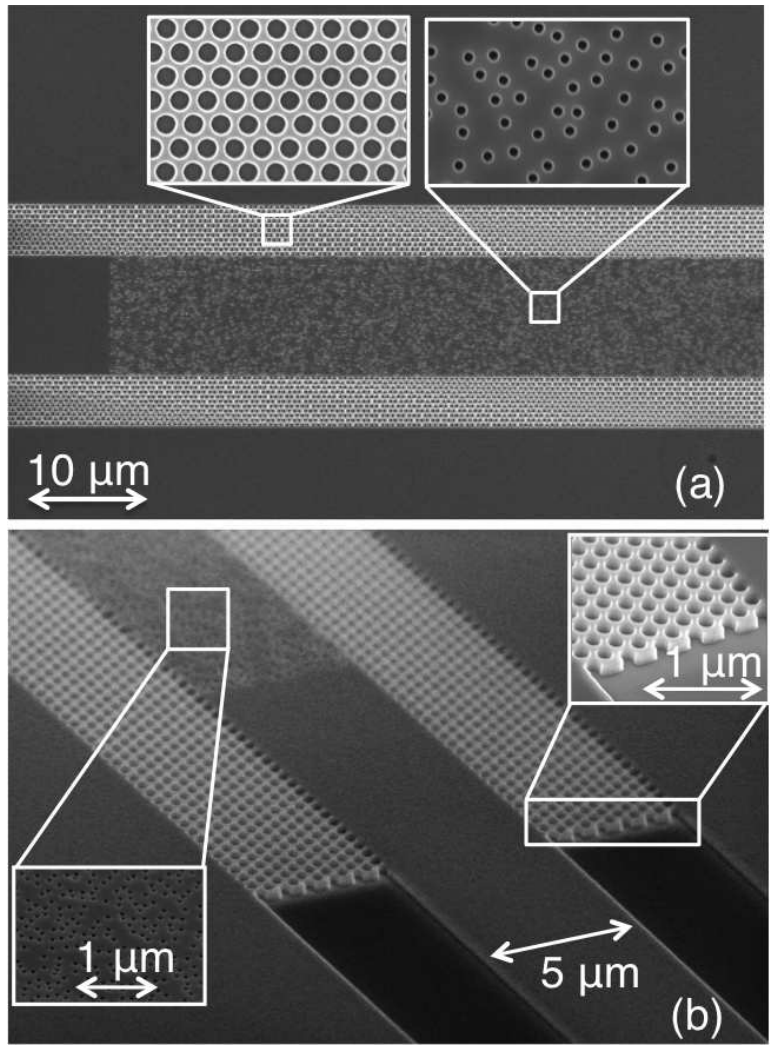}}
\caption{
\label{fig:sem} 
{\bf Experimental realization of disordered waveguides.}
Top-view {\bf a} and tilt-view {\bf b} scanning electron microscopy (SEM) images of an optical waveguide fabricated in a silicon membrane on top of silica. The two sidewalls of the waveguide consist of triangular lattices of air holes (lattice constant 440 nm, hole radius 154 nm). They possess a 2D photonic bandgap and behave like reflecting walls for light incident from all angles in the waveguide. The probe light is coupled from a silicon ridge waveguide to an empty photonic crystal waveguide, then impinge onto a random array of air holes (hole diameter 100 nm, and areal density 6 \%) inside the waveguide.}
\end{figure}

We designed and fabricated two-dimensional (2D) disordered waveguide structures in a 220~nm silicon layer on top of 3~$\mu$m buried oxide. The patterns were written by electron beam lithography and etched in an inductively coupled plasma reactive ion etcher. As shown in the scanning electron microscope images in Fig.~\ref{fig:sem}, the waveguide has sidewalls made of periodic arrays of air holes. They possess a 2D photonic bandgap that covers the wavelength range of the probe light ($\lambda$ = 1500~nm -- 1520~nm), thus providing optical confinement in the plane of the waveguide. Light enters the waveguide from an open end and is incident onto a 2D array of air holes inside the waveguide. The random pattern of air holes causes light to scatter while going through the waveguide. The transport mean free path $\ell$ is determined by the size and density of air holes. Light localization will occur if the length of the random array $L$ exceeds the localization length  $\xi=(\pi/2)N\ell$, where $N= 2W/(\lambda/n_e)$ is the number of propagating modes in the waveguide, $W$ is the waveguide width, $\lambda$ is the optical wavelength in vacuum, and $n_e$ is the effective index of refraction of the random medium. Since $N$ scales linearly with $W$, $\xi$ can be easily tuned by varying the waveguide width. Therefore, by changing the waveguide geometry ($L$, $W$), we can reach both the diffusion regime ($\ell < L < \xi$) and localization regime ($L > \xi$)\cite{1997_Beenakker,1998_Brouwer}. Although there is no mobility edge\cite{2013_Maret_3D_localization} in such a system, it is not essential for our goal of observing position-dependent diffusion. 

\begin{figure}
\centering{\includegraphics[width=6in,angle=-0]{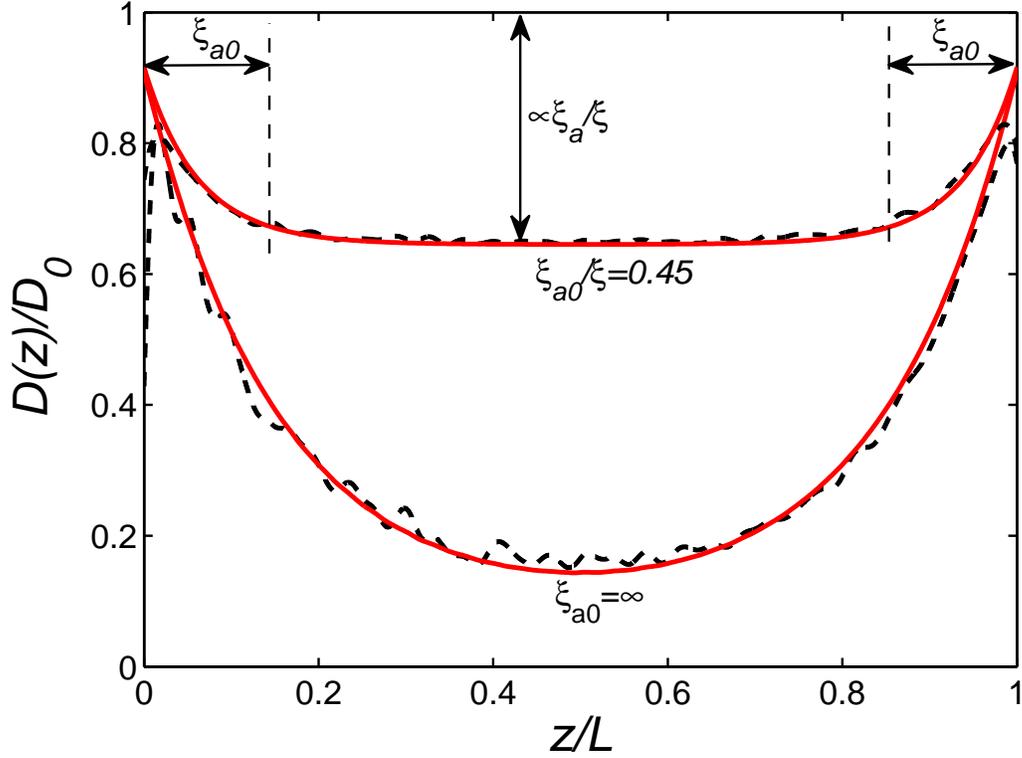}}
\caption{
\label{fig:simulations} 
{\bf Numerical simulations of position-dependent diffusion.}
Numerically calculated diffusion coefficient $D(z)$ (dashed curves) at position $z$ ($0<z<L$) in a waveguide filled with randomly positioned scatterers. The waveguide has length $L=3\xi$ ($\xi$ is the localization length) and supports $N=10$ modes. The solid curves represent the prediction of the self-consistent theory of localization. $D_0$ denotes the diffusion coefficient that ignores interference effects.  In the absence of absorption (diffusive absorption length $\xi_{a0}=\infty$), $D(z)$ drops to a minimum of $0.17D_0$ in the middle of the waveguide. With the addition of absorption ($\xi_{a0}/\xi=0.45$), $D(z)$ exhibits a plateau for $\xi_{a0}<z<L-\xi_{a0}$, and its value $D_p$ is determined by the ratio $\xi_{a0}/\xi$. }
\end{figure}

In order to apply the self-consistent theory of localization to the analysis of the experimental data below, we first validate it with numerical simulations under conditions close to those in the experiment. We computed the position-dependent diffusion coefficient without making any assumption about the nature or strength of wave interference (see the Methods section). Figure~\ref{fig:simulations} plots the calculated $D(z)/D_0$ (the lower dashed line) for $L/\xi = 3.0$, where $D_0$ is the diffusion coefficient without renormalization.. The $z$ axis is parallel to the waveguide, and the random array extends from $z=0$ to $z= L$. The renormalized $D(z)$ drops to $0.17D_0$ in the middle of the random waveguide ($z= L/2$). Using the self-consistent theory of localization (see the Supplementary Information) we calculate $D(z)/D_0$ (the lower solid line in Fig.~\ref{fig:simulations}) and it is in excellent agreement with the ab-initio simulation without any fitting parameters. Previous studies show that further into the localization regime where resonant tunneling dominates wave transport, the self-consistent theory of localization underestimates the energy density inside the random system that is strongly affected by the presence of necklace states\cite{2010_Tian_PRL}. In our experiment we avoid such a regime and stay where the self-consistent theory of localization holds\cite{2013_Yamilov_Localization_with_Absorption}.  

Another factor we shall consider is the dissipation of light in the random waveguide. Within the wavelength range of the probe, light absorption by silicon or silica is negligible. However, light is scattered out of the waveguide plane by the random array of air holes. Such scattering allows us to monitor the intensity distribution inside the system from the vertical direction. The question is whether the out-of-plane scattering can be treated as incoherent dissipation. In a periodic array of scatterers, long-range correlation of light fields makes the waves scattered from different locations phase coherent and their interference in the far field zone determines the out-of-plane leakage. In contrast, in a random array of scatterers, the fields are correlated\cite{1986_Shapiro_C1_correlations,1988_Feng} only within a distance of the order one transport mean free path $\ell$, and waves from different coherent regions of size $\ell\times\ell$ have uncorrelated phases. Since there are a large number of such coherence regions $\ell\times\ell$ in our waveguides $W\times L$, the overall leakage may be considered incoherent and treated effectively as absorption. 

To illustrate the effect of dissipation on position-dependent diffusion, we perform numerical simulations. The diffusive absorption length in the random system is $\xi_{a0} = \sqrt{D_0 \tau_a}$, where $\tau_a$ is the ballistic absorption time. When $\xi_{a0}$ becomes smaller than the localization length $\xi$, the effect of dissipation is significant. Figure~\ref{fig:simulations} plots the calculated $D(z)/D_0$ in the random waveguide with $\xi_{a0}/\xi = 0.45$ (the upper dashed line) in comparison to that with $\xi_{a0}=\infty$ (no absorption).  The suppression of diffusion is weakened by the absorption, and a plateau for the renormalized diffusion coefficient is developed inside the disordered system. This result can be understood as follows. Dissipation  suppresses feedback from long propagation paths, limiting the effective size of the system\cite{1998_Brouwer} to the order of diffusive absorption length for any position that is more than one $\xi_{a0}$ away from the open boundary ($\xi_{a0}<z<L-\xi_{a0}$)\cite{2013_Yamilov_Localization_with_Absorption}. Thus the renormalized $D$ reaches a constant value equal to that of an open system of dimension $\sim 2 \xi_{a} = 2 \sqrt{D \tau_a}$. In the remaining regions that are within one $\xi_{a0}$ to the boundary  ($z<\xi_{a0}$ and $L-z<\xi_{a0}$), the diffusion coefficient is still position dependent due to leakage through the boundary and $D$ increases toward the value of $D_0$ without renormalization. We note that the extent of these regions $\xi_{a0}$ is much greater than the transport mean free path $\ell$. The latter determines the boundary region where the diffusion approximation is not accurate even without wave interference\cite{2007_Akkermans_book}. Figure~\ref{fig:simulations} also shows the prediction of self-consistent theory of localization in the presence of dissipation (the upper solid line), and it agrees well with the numerical result.  

Experimentally there are three advantages to using the planar waveguide geometry. First, it allows a precise fabrication of the desired system using lithography so that the parameters such as the transport mean free path can be accurately controlled. Second, the localization length $\xi\propto W$ can be varied by changing the waveguide width, while the diffusive absorption length $\xi_{a0}$ remains fixed. This allows us to separate the effects of localization and dissipation by testing waveguides of different width. Finally, unlike 2D random systems\cite{2011_Wiersma}, the additional confinement of light by the waveguide sidewalls makes $\xi$ scale linearly with $\ell$. Even if scattering is relatively weak ($k \ell \gg 1$, where $k$ is the wavenumber), the waveguide length $L$ can easily exceed $\xi$ so that the localization effect is strong enough to modify diffusion. Instead of designing the disorder to maximize scattering (minimizing $k \ell$), we deliberately lower the density of air holes to mitigate the out-of-plane scattering loss and maximize the ratio $\xi_{a0}/\xi$.

\begin{figure}
\centering{\includegraphics[width=5in,angle=-0]{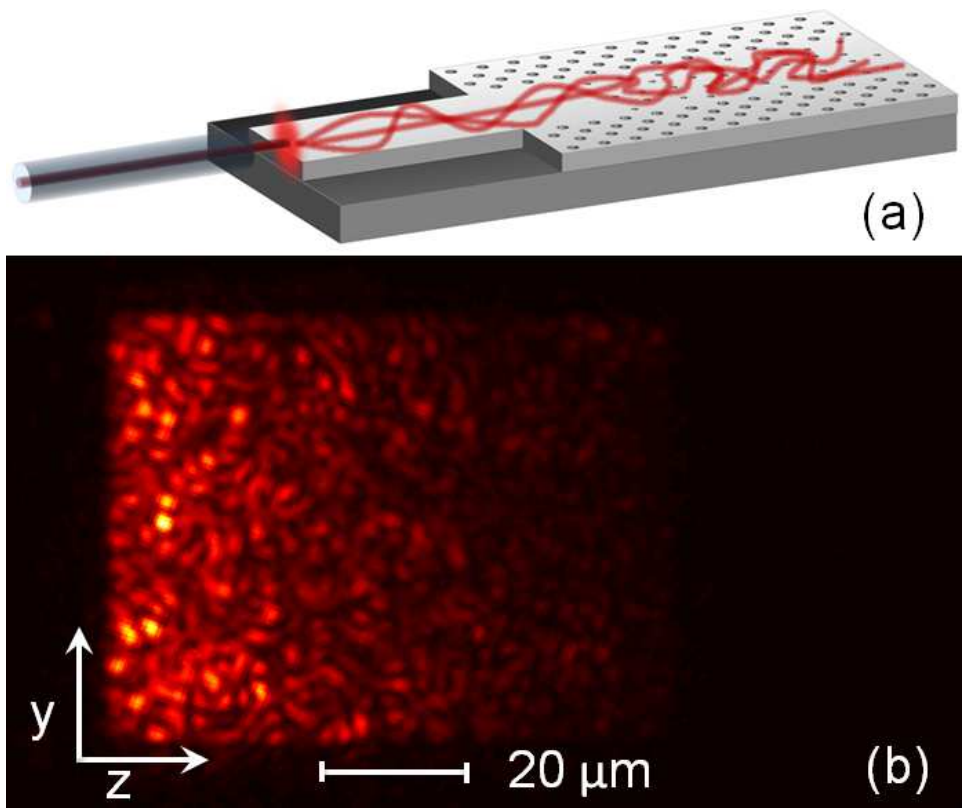}}
\caption{
\label{fig:setup} 
{\bf Optical measurement of light inside random medium.} 
{\bf a}, A schematic depiction of light coupling from a lensed fiber to a waveguide in the silicon membrane. 
{\bf b}, A near-field optical image showing the spatial distribution of light intensity in a random waveguide with a configuration similar to that shown in Fig.~\ref{fig:sem}. The wavelength of the probe light is 1510~nm. A 50$\times$ objective lens collects the light scattered by the air holes out of the waveguide plane and images onto a camera. }
\end{figure}

In the optical measurement, output from a wavelength-tunable continuous-wave laser (HP 8168F) was coupled to the waveguide through a single-mode polarization-maintaining lensed fiber as shown schematically in Fig.~\ref{fig:setup}a. To ensure efficient confinement in disordered waveguide, the transverse-electric (TE) polarization (electric field in the plane of the waveguide) of the incident light was chosen. A near-field optical image of the spatial distribution of light intensity across the structure surface was taken by collecting light scattered out of plane using a 50X objective lens (numerical aperture 0.42) and recorded by an InGaAs camera (Xenics Xeva 1.7-320). An example of the near-field image is shown in Fig.~\ref{fig:setup}b, from which the intensity distribution $I(y,z)$ is extracted. The $y$ axis is in the waveguide plane and perpendicular to the direction of incident beam. $I(y,z)$ was integrated over the cross section of the waveguide (along the $y$ axis) to obtain the evolution of intensity $I(z)$ in the incident direction (parallel to the $z$ axis). For each configuration (width $W$, length $L$, transport mean free path $\ell$) of the disordered waveguides, $I(z)$ was averaged over two random realizations of air holes and fifty input wavelengths equally spaced between 1500~nm and 1520~nm. The wavelength spacing was chosen to produce independent intensity distributions.

\begin{figure}
\vskip -1in
\centering{\includegraphics[width=6in,angle=-0]{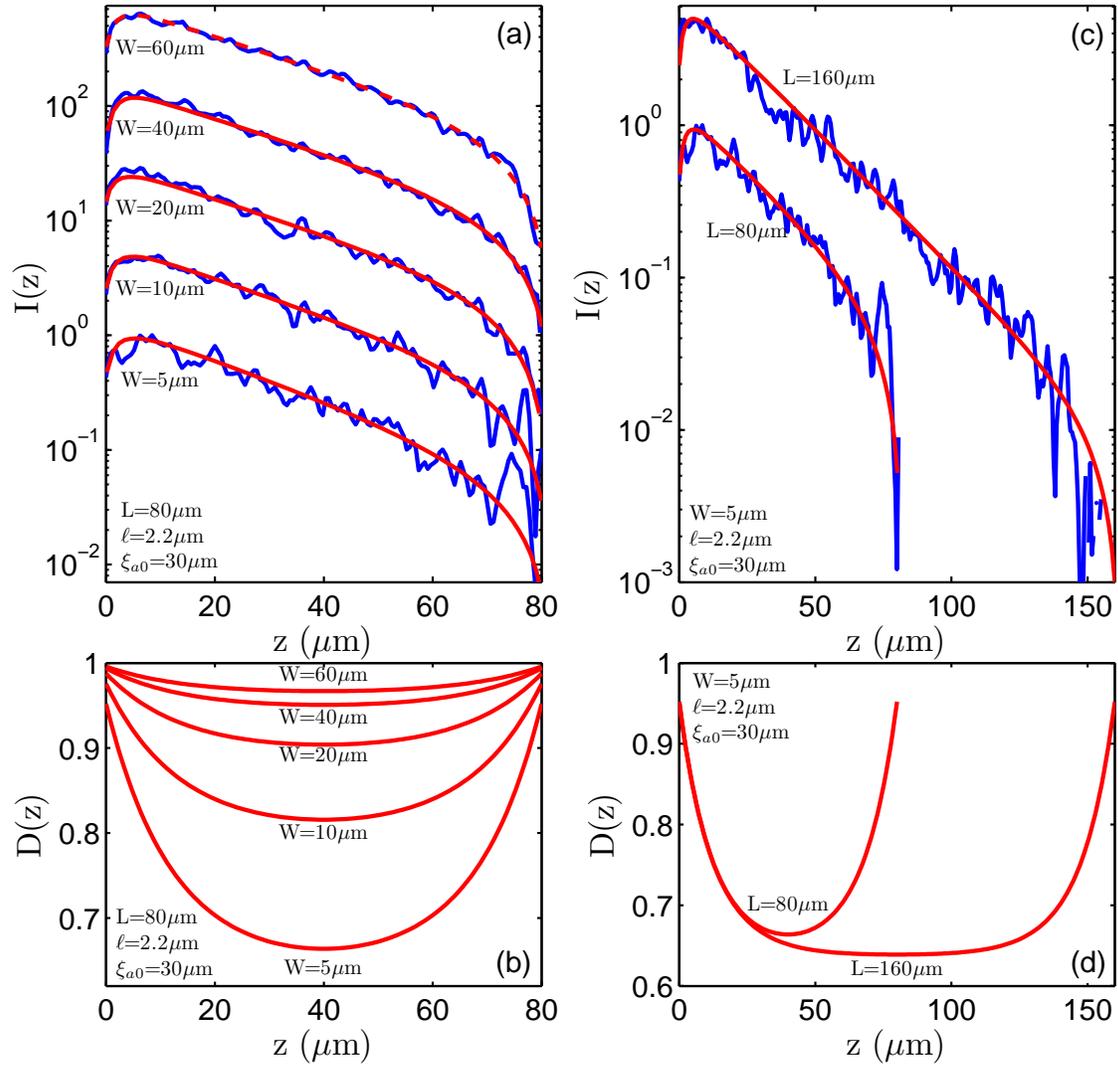}}
\vskip -0.5in
\caption{\label{fig:weak_scattering} 
{\bf Experimental measurement of light transport inside disordered waveguides.}
{\bf a},~Experimentally measured light intensity $I(z)$ inside random waveguides of different width $W$ and constant waveguide length $L$=80~$\mu$m (blue solid lines). The curves are vertically shifted for a clear view. $\ell$=2.2~$\mu$m and $\xi_{a0}$=30~$\mu$m are found by fitting the $W$=60~$\mu$m sample with the self-consistent theory of localization (red dashed line). With these parameters, the self-consistent theory of localization predicts $I(z)$ for other samples of $W$=40~$\mu$m, 20~$\mu$m, 10~$\mu$m, 5~$\mu$m  (red solid curves), and the prediction agrees well with the experimental data without any free parameter except the vertical intensity scale. 
{\bf b},~Position-dependent diffusion coefficients for the five samples in {\bf a}.
{\bf c},~Experimentally measured $I(z)$ of two waveguides with the same width $W$=5~$\mu$m but different length, $L$=80~$\mu$m, 160~$\mu$m (blue solid curves). Red solid curves represent the prediction of the self-consistent theory of localization using the same values of $\ell$ and $\xi_{a0}$ as in {\bf a}. 
{\bf d}, Diffusion coefficients $D(z)$ for the two samples in {\bf c}, showing the saturation of $D$ inside the longer sample $L$=160~$\mu$m. }
\end{figure}

Figure~\ref{fig:weak_scattering}a shows the measured $I(z)$ inside the ensembles of random waveguides of $W$ varying from 60~$\mu$m to 5~$\mu$m (blue solid lines). All other parameters are kept the same, $L$ is fixed at 80~$\mu$m, the diameter of air holes is 100~nm, and average (center-to-center) distance of adjacent holes is 390~nm.  $\ell$ and $\xi_{a0}$ are obtained by fitting the least localized sample, $W$=60~$\mu$m (longest $\xi$), with the self-consistent theory of localization (red dashed line). We find that $\ell = 2.2 \pm 0.1$~$\mu$m and $\xi_{a0}=30\pm 0.5$~$\mu$m. With the parameters found from the $W$=60~$\mu$m sample, the self-consistent theory of localization successfully predicts the decay for $I(z)$ in all other samples with $W$=40~$\mu$m, 20~$\mu$m, 10~$\mu$m, 5~$\mu$m (red solid lines). We stress that the excellent agreement with the experimental data are obtained without any free parameter except the vertical intensity scale. The position-dependent diffusion coefficients $D(z)$ corresponding to the red curves in Fig.~\ref{fig:weak_scattering}a are shown in Fig.~\ref{fig:weak_scattering}b. We can clearly see that the diffusion coefficient is reduced inside the sample, and its value varies along $z$. Farther away from the open boundary, $D$ has a smaller value. In the narrower waveguides, the reduction of $D$ is larger due to stronger localization effect. In the most localized sample of $W$ = 5$\mu$m, $D$ is reduced to $0.65 D_0$ at $z = L/2$. In an attempt to further reduce $D$, we double the length of random system $L$ to 160~$\mu$m. As shown in Fig.~\ref{fig:weak_scattering}c,d for $W$=5~$\mu$m, the minimal $D$ no longer decreases, instead it saturates in the middle of the random waveguide. This behavior is attributed to dissipation which suppresses localization. As the system length $L$ becomes much larger than the diffusive absorption length $\xi_{a0}$, $D(z)$ saturates to a constant value $D_p$ inside the disordered waveguide, similar to the simulation result shown in Fig.~\ref{fig:simulations}. 

\begin{figure}
\vskip -0.5in
\centering{\includegraphics[width=3in,angle=-0]{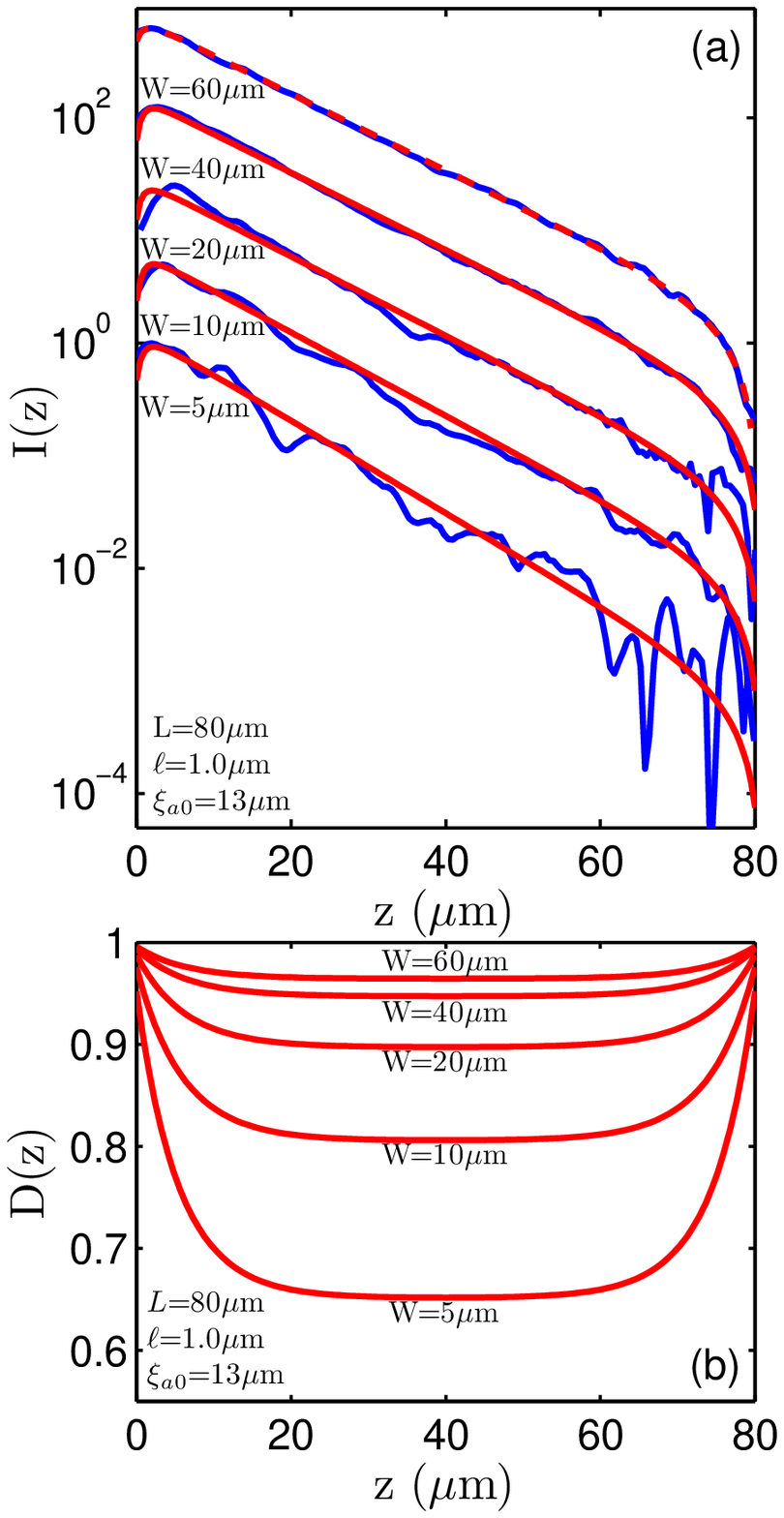}}
\vskip -0.5in
\caption{
\label{fig:strong_scattering} 
{\bf Tuning the diffusion coefficient via the interplay of localization and dissipation.}
{\bf a},~Experimentally measured light intensity $I(z)$ inside random waveguides in the deep saturation regime $\xi_{a0}\ll L$ (blue solid lines). The curves are vertically offset for a clear view.  The length and width of the waveguides are given in the graph. $\ell=1.0$~$\mu$m and $\xi_{a0}=13$~$\mu$m are found by fitting the $W$=60~$\mu$m sample with the self-consistent theory of localization (red dashed line). This is then used to predict $I(z)$ for other samples $W$=40~$\mu$m, 20~$\mu$m, 10~$\mu$m, 5~$\mu$m (red solid curves). The prediction of the self-consistent theory of localization is in good agreement with the experimental data with no fitting parameters except the vertical intensity scale.
{\bf b},~Diffusion coefficients $D(z)$ for all samples in {\bf a} are saturated in the region $\xi_{a0}<z<L-\xi_{a0}$. The saturated values are smaller for narrower waveguides. }
\end{figure}

Finally we exploit the interplay between dissipation and localization to tune the saturated value of the diffusion coefficient inside the random system. To this end, we increase the density of scatterers to reach the deep saturation region $\xi_{a0}\ll L$. In the second set of samples, the diameter of air holes is 150~nm, and the average distance between adjacent holes is 370~nm. Waveguide length $L$ is set at 80~$\mu$m and $W$ varies from 5~$\mu$m to 60~$\mu$m. Experimental data of measured intensity $I(z)$ inside the random waveguides are presented in Fig.~\ref{fig:strong_scattering}a. By fitting the $W$=60~$\mu$m sample with the self-consistent theory of localization (red dashed line), we obtain $\ell=1.0\pm 0.1$~$\mu$m and $\xi_{a0}=13.0\pm 0.3$~$\mu$m. With these values, the self-consistent theory of localization gives the spatial profiles of $I(z)$ for $W$=40~$\mu$m, 20~$\mu$m, 10~$\mu$m, 5~$\mu$m (red solid lines), which agree well with the experimental data with no fit. The corresponding values of $D(z)$ are plotted in Fig.~\ref{fig:strong_scattering}b. Due to stronger scattering (smaller $\ell$) and larger out-of-plane loss (shorter $\xi_{a0}$), $D(z)$ for all five samples displays a well-developed plateau inside the sample. The  saturated value $D_p$ decreases with $W$ -- the narrower waveguide has smaller $D_p$. Hence, without changing the disorder or altering the dissipation rate, we can control the diffusion inside a random system by merely varying its geometry ($W$ in this case). 

Renormalization of the diffusion coefficient $D$ has long been considered as a {\it theoretical} approach to tackle the problem of localization. Here we present direct experimental evidence of suppressed diffusion of light {\it inside} the random systems. This is an intrinsic wave phenomenon, hence our conclusions also apply to  acoustic wave, microwave and even the de Broglie wave of electrons. The diffusion coefficient $D$ is altered by wave interference and such modification depends on the position inside an open random system. By varying the size and shape of the random system, we are able to control the strength of wave interference and consequently the degree of renormalization of $D$. We also show that the presence of dissipation prevents $D$ from approaching zero and sets a limit for the minimal $D$ that can be reached by the localization effect. Such effect of dissipation is expected to be similar to that of dephasing in the electronic systems\cite{1991_Altshuler}. The interplay between localization and dissipation enables us to tune the value of $D$ inside random systems. The results presented in this work are obtained by directly probing light transport inside the random media, which allows us to extract the diffusion coefficient anywhere inside the system. Such experiments open the path to measure other transport properties inside random systems, e.g., intensity correlations and fluctuations.

\begin{methods}

\noindent {\it Numerical simulations}:
The numerical data shown in Fig.~\ref{fig:simulations} were obtained from the ab-initio simulation of a  monochromatic scalar wave propagating in a 2D waveguide filled with random scatterers. With continuous-wave excitation from one end of the waveguide ($z=0$), we computed the energy density ${\cal W}(z)$ and flux along the $z$ axis $J_z(z)$, and averaged them over the cross section of the waveguide. The position-dependent diffusion coefficient was found from Fick's law as $D(z) = -\langle J_z(z)\rangle / \left[ d\langle{\cal W}(z)\rangle/d z\right]$, where angular brackets denote the ensemble average. Using a supercomputer, we numerically simulated an ensemble of $10^6$ waveguides with different disorder configurations. To obtain the value of the diffusion coefficient $D_0$ without renormalization, we used the single parameter scaling and the expression for direction-resolved flux, as detailed in Supplementary Information. In the absorbing samples we used the continuity equation to calculate the absorption time $\tau_a$.

\noindent {\it Self-consistent theory of localization}:
The application of self-consistent theory of localization with a position-dependent diffusion coefficient to  disordered waveguides was described in Ref.~\cite{2010_Payne_PRL}. It involves the diffusion equation which defines the return probability and a self-consistency equation that relates diffusion coefficient $D(z)$ to the return probability. We solved these two equations by iteration until we found $D(z)$ which satisfied both equations, see the Supplementary Information for more details. 

\noindent {\it Design of photonic crystal walls for 2D waveguides}:
The triangular lattice of air holes that form the sidewalls of the random waveguide were designed to have a 2D photonic bandgap for TE polarized light in the wavelength range of 1450~nm -- 1550~nm. The photonic band structure was calculated with the plane wave expansion method\cite{2001_Johnson_Joannopoulos_mpb}.

\end{methods}

\section*{References}
%\bibliography{../../Bibliography/latex_bibliography}

\begin{addendum}
 \item We are indebted to Patrick Sebbah and Shivakiran Bhaktha for their insight in selecting the experimental geometry. We acknowledge Seng Fatt Liew, Douglas Stone, Arthur Goetschy, Boris Shapiro, Chushun Tian and Sergey Skipetrov for useful discussions. We also thank Michael Rooks for suggestions regarding sample fabrication. This work was supported by National Science Foundation under grants Nos. DMR-1205307, DMR-1205223 and ECCS-1128542. Computational resources were provided under the Extreme Science and Engineering Discovery Environment (XSEDE) grant No. DMR-100030. Facilities use was supported by YINQE and NSF MRSEC DMR-1119826.
 \item[Author contributions] A.Y. and H.C. initiated the study and designed the experiments. R.S. and B.R. fabricated the samples. H.N. designed the waveguide walls. R.S. collected the experimental data. R.S. and A.Y. analyzed the data. B.P. and A.Y. performed ab-initio numerical simulation. A.Y. and H.C. prepared the manuscript.
 \item[Additional information] Supplementary information is available in the online version of the paper. Reprints and permissions information is available online at www.nature.com/reprints. Correspondence and requests for materials should be addressed to A.Y. and H.C.
 \item[Competing Interests] The authors declare that they have no competing financial interests.
\end{addendum}

\newpage
\setcounter{figure}{0}
\setcounter{page}{1}
\renewcommand{\figurename}{Figure S}

\begin{center}
{\bf SUPPLEMENTARY INFORMATION}\\ 
\vskip 1cm
{\bf Position-dependent diffusion of light in disordered waveguides}\\
Alexey G. Yamilov, Raktim Sarma, Brandon Redding, Ben Payne, Heeso Noh \& Hui Cao
\end{center}
\vskip 1cm

\section{Calculation of position-dependent diffusion coefficient $D(z)$} 

In the ab-initio numerical simulation, we consider a monochromatic scalar wave $E(\mathbf{r})e^{-i \omega t}$ propagating in a 2D volume-disordered waveguide of width $W$ and length $L \gg W$. The wave field $E(\mathbf{r})$ obeys the 2D Helmholtz equation:
\begin{equation}
\left\{\nabla^2 + k^2\left[1 + \delta\epsilon(\mathbf{r}) \right]\right\} E(\mathbf{r}) = 0.
\label{eq:helmholtz}
\end{equation}
Here $k=\omega/c$ is the wavenumber and $\delta\epsilon(\mathbf{r})=(1+i\alpha)\delta\epsilon_r(\mathbf{r})$, where $\delta\epsilon_r(\mathbf{r})$ describes the random fluctuation of the dielectric constant, and $\alpha>0$ denotes the strength of absorption. The system is excited from one open end ($z=0$) of the waveguide (extending from $z=0$ to $z=L$) by illuminating each of the guided modes with a unit flux. The wave field $E(\mathbf{r})$ throughout the random medium is computed with the transfer matrix method for a given realization of disorder\cite{2010_Payne_PRL}. From $E(\mathbf{r})$ we calculate the energy density ${\cal W}(z)$ and the flux $J_z(z)$ along the $z$ axis (parallel to the waveguide axis). These two quantities are averaged over the cross section of the waveguide at each $z$ and give the diffusion coefficient:
\begin{equation}
D(z) = -\langle J_z(z)\rangle / \left[ d\langle{\cal W}(z)\rangle/d z\right],
\label{eq:Dofz_definition}
\end{equation}
where the averages $\langle \ldots \rangle$ are taken over a statistical ensemble of $10^6$ disorder realizations. 

In order to compare our numerical results for $D(z)$ with the self-consistent theory of localization, we need to have the value of the diffusion coefficient without renormalization due to the wave interference effects $D_0=v\ell/2$. To estimate the transport mean free path $\ell$ in our model we perform a set of simulations for different waveguide lengths $L$, exploring both the regime of diffusion $L<\xi$ and that of Anderson localization $L>\xi$. We computed numerically the conductance $g$ as the sum of transmission coefficients from all incoming to all outgoing waveguide modes. The dependencies of the average $\langle g \rangle$ and variance $\mathrm{var}(g)$ on $L$ are fitted by the analytical expressions obtained by Mirlin in Ref.~\cite{2000_Mirlin} using the supersymmetry approach with $\ell$ being the only fit parameter. To find the diffusive speed $v$ we use the definition of diffusive flux in the forward ($+z$) direction $J^{(+)}_z(z)$ and the backward ($-z$) direction $J^{(-)}_z(z)$ with respect to the  propagation direction\cite{1999_van_Rossum} 
\begin{equation}
\langle J^{(\pm)}_z(z)\rangle = (v/\pi)\langle{\cal W}(z)\rangle \mp (D(z)/2)d\langle {\cal W}(z)\rangle/dz.
\end{equation}
Combining the two components we find the diffusive speed
\begin{equation}
v=2\left( \langle J^{(+)}_z(z)\rangle + \langle J^{(-)}_z(z)\rangle\right)/\langle{\cal W}(z)\rangle.
\label{eq:v}
\end{equation}
Dashed lines in Fig.~\ref{fig:simulations} depict $D(z)$ found in equation~(\ref{eq:Dofz_definition}) normalized by $D_0$. 

In the dissipative random waveguides, the characteristic absorption time $\tau_a$ is determined numerically using the condition of flux continuity $d\left\langle J_z(z)\right\rangle/dz = (1/\tau_a) \left\langle{\cal W}(z)\right\rangle$.  The desired diffusive absorption length $\xi_{a0}=\sqrt{D_0\tau_a}$ can be obtained by the proper choice of $\alpha$ in equation~(\ref{eq:helmholtz}).

\section{Self-consistent theory of localization}

%The question of how the diffusion coefficient evolves (scales) from the unrenormalized value of $D_0$ toward the limit of $D=0$ has been addressed in Ref.~\cite{2000_van_Tiggelen,2008_Cherroret} by extending the original self-consistent theory of Vollhardt and W\"olfle\cite{1980_Vollhardt_Wolfle,1993_Kroha_self_consistent} to systems of finite size. The key prediction of the modified self-consistent theory (SCT) is that the diffusion coefficient is no longer a constant but varies spatially. This is because the return probability responsible for the renormalization of diffusion is position-dependent in a system of finite size. As the wave explores the larger and larger neighborhood of a source point, the return probability is a fundamentally nonlocal quantity, sensitive to the proximity of a boundary.

The self-consistent theory starts with the Green's function $G(\mathbf{r},\mathbf{r}')$ of equation~(\ref{eq:helmholtz}) with $\delta\epsilon(\mathbf{r})=\delta\epsilon_r(\mathbf{r})+i\alpha$. In a random waveguide, the disorder-averaged function ${\hat C}(\mathbf{r}, \mathbf{r}')=(4\pi W D_0/cL)\langle \left| G(\mathbf{r}, \mathbf{r}') \right|^2 \rangle$ obeys self-consistent equations in a dimensionless form\cite{2008_Cherroret,2010_Payne_PRL}: 
\begin{eqnarray}
&&\left[\left(\frac{L}{\xi_{a0}}\right)^2 - \frac{\partial}{\partial \zeta} d(\zeta)
 \frac{\partial}{\partial \zeta} \right] {\hat C}(\zeta,\zeta')
= \delta(\zeta-\zeta'),
\label{eq:sceq1}
\\
&&\frac{1}{d(\zeta)} =  1+\frac{2L}{\xi}
{\hat C}(\zeta,\zeta),
\label{eq:sceq2}
\end{eqnarray}
where $d(\zeta) = D(\zeta)/D_0$ and all position-dependent quantities are functions of the longitudinal coordinate $\zeta = z/L$. The quantity ${\hat C}(\zeta,\zeta)$, which renormalizes the diffusion coefficient, is proportional to the return probability at $\zeta$. Assuming first that $d(\zeta)\equiv 1$, equations~(\ref{eq:sceq1},\ref{eq:sceq2}) are solved by iteration with the  boundary conditions:
\begin{eqnarray}
{\hat C}(\zeta,\zeta^{\prime}) \mp
\frac{z_0}{L} d(\zeta) \frac{\partial}{\partial \zeta}
{\hat C}(\zeta,\zeta^{\prime}) = 0
\label{eq:bc}
\end{eqnarray}
at $\zeta = 0$ and $\zeta = 1$. The $z_0=(\pi/4)\ell$ is the so-called extrapolation length\cite{1999_van_Rossum}. 

After the self-consistent solution of equations~(\ref{eq:sceq1}-\ref{eq:bc}) has been found, we find the intensity distribution inside the sample by replacing the delta-function source in equation~(\ref{eq:sceq1}) with $\left(L/\ell\right)\exp\left[-\zeta/(\ell/L)\right]$. This source term represents the exponential attenuation of the incident ballistic signal.

\section{Experimental details}

The experimental setup for optical characterization is shown in Fig.~S1(a). We used a single-mode polarization-maintaining fiber to deliver the probe light into a silicon ridge waveguide on a SOI substrate. The fiber was tapered at the end to focus the laser beam to a spot of diameter $\sim 2.5~\mu$m at the edge of the wafer. The ridge waveguide had the same width as the random waveguide it was connected to, which varied from 5 micron to 60 micron [Fig.~S1(b)]. However, the height of the silicon waveguide was merely 220~nm, so some of the input light did not couple into the waveguide; instead it propagated above or below the waveguide. To avoid such stray light, the ridge waveguide was tilted by 30~degrees with respect to the incident direction of the light from the fiber (approximately normal to the edge of the wafer). The ridge waveguide was made 2.5~mm long, so that the random waveguide structure is far from the direct path of the stray light. In addition, uniform illumination of the front surface of the random structure inside the waveguide was ensured by positioning the tapered fiber approximately at the center of the input facet of the ridge waveguide. The spatial distribution of light intensity over the sample was imaged by an objective lens onto an IR CCD camera [not shown in Fig.~S1(a)]. 

%%%%%%%%%%%%%%%%%%%%%%%%%%%%%%%%%%%%%%%%%%%%%%%%%%%%%%%%%%%%%%%%%%%%%%%%%%%%%%%
\begin{figure}
\centering{\includegraphics[width=6.5in,angle=-0]{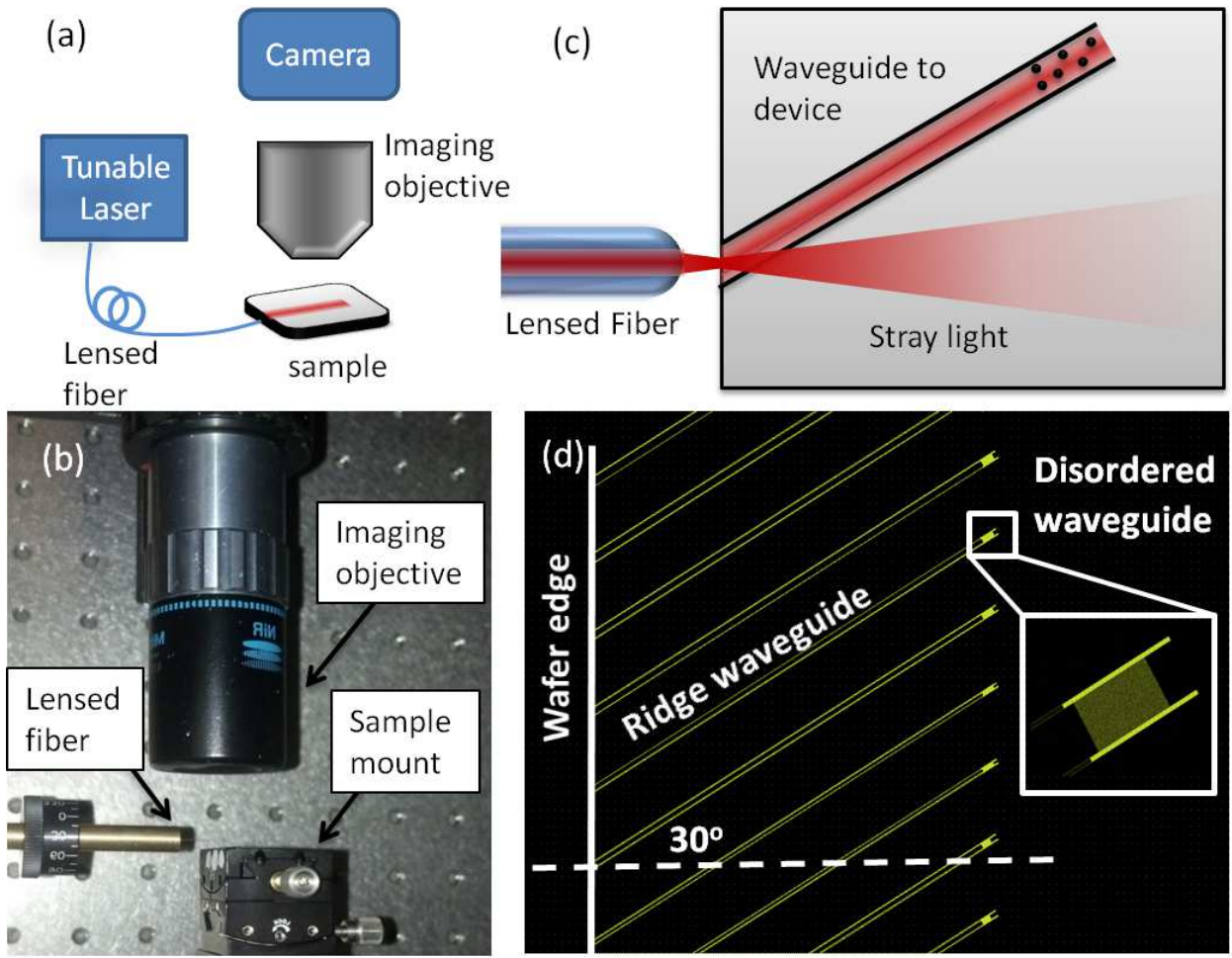}}
\caption{{\bf Optical measurement setup.}
{\bf a},~Schematic of experimental setup for measuring light transport inside the random waveguide. 
{\bf b},~Photograph of the experimental setup.
{\bf c},~Schematic of the sample layout showing the ridge waveguides coupling the probe light from the edge of the wafer to the random waveguides with photonic crystal sidewalls.
{\bf d},~Layout of the fabricated structures studied experimentally.}
\end{figure}
%%%%%%%%%%%%%%%%%%%%%%%%%%%%%%%%%%%%%%%%%%%%%%%%%%%%%%%%%%%%%%%%%%%%%%%%%%%%%%%

\end{document}